\documentclass[zpreprint,zbstdefault]{zeus_paper}
\usepackage[english]{babel}
\newcommand{\Zdetdesc}{%
A detailed description of the ZEUS detector can be found 
elsewhere~\cite{zeus:1993:bluebook}. A brief outline of the 
components that are most relevant for this analysis is given
below.\xspace}

\chardef\usc=95
\chardef\til=126
\catcode`\@=11 
\DeclareRobustCommand\xdotspace{\futurelet\@let@token\@xdotspace}
\def\@xdotspace{%
  \ifx\@let@token.\else
  \ifx\@let@token\bgroup.\else
  \ifx\@let@token\egroup.\else
  \ifx\@let@token\/.\else
  \ifx\@let@token\ .\else
  \ifx\@let@token~.\else
  \ifx\@let@token!.\else
  \ifx\@let@token,.\else
  \ifx\@let@token:.\else
  \ifx\@let@token;.\else
  \ifx\@let@token?.\else
  \ifx\@let@token/.\else
  \ifx\@let@token'.\else
  \ifx\@let@token).\else
  \ifx\@let@token-.\else
  \ifx\@let@token\@xobeysp.\else
  \ifx\@let@token\space.\else
  \ifx\@let@token\@sptoken.\else
   .\space
   \fi\fi\fi\fi\fi\fi\fi\fi\fi\fi\fi\fi\fi\fi\fi\fi\fi\fi}
\catcode`\@=12 

\newcommand{\stru}[2]{%
   \relax\ifmmode\hbox{\vrule height#1 depth#2 width0pt}%
   \else\vrule height#1 depth#2 width0pt\fi}

\newcommand{\Ronum}[1]{\uppercase\expandafter{\romannumeral#1}}
\newcommand{\ronum}[1]{\expandafter{\romannumeral#1}}
\DeclareRobustCommand{\LaTeXZ}{%
  \LaTeX\kern-.05em4\kern-.1em
  {\raisebox{-0.2ex}{$\scriptstyle\text{ZEUS}$}}\xspace}

\DeclareMathAlphabet{\mathbf}{OT1}{cmr}{bx}{sl}
\newcommand{\eVdist}{\kern-0.06667em}

\newcommand{\Gev}{{\text{Ge}\eVdist\text{V\/}}}

\newcommand{\mev}{{\,\text{Me}\eVdist\text{V\/}}}
\newcommand{\gev}{{\,\text{Ge}\eVdist\text{V\/}}}

\newcommand{\pbi}{\,\text{pb}^{-1}}
\newcommand{\fbi}{\,\text{fb}^{-1}}

\newcommand{\Tesla}{\,\text{T}}

\newcommand{\slashfrac}[2]{%
  \raisebox{0.5ex}{\ensuremath #1}\kern-0.12em/\kern-0.08em
  \raisebox{-.8ex}{\ensuremath #2}}

\newcommand{\sqr}[3]{%
    {\vcenter{\hrule height.#3ex\hbox{\vrule width.#2ex height#1ex
     \kern#1ex\vrule width.#3ex}\hrule height.#2ex}}}

\catcode`\@=11 
\newcommand{\parenbar}{\mathpalette\p@renb@r}
\def\p@renb@r#1#2{\vbox{%
  \ifx#1\scriptscriptstyle \dimen@.7em\dimen@ii.2em\else
  \ifx#1\scriptstyle \dimen@.8em\dimen@ii.25em\else
  \dimen@1em\dimen@ii.4em\fi\fi \offinterlineskip
  \ialign{\hfill##\hfill\cr
    \vbox{\hrule width\dimen@ii}\cr
    \noalign{\vskip-.3ex}%
    \hbox to\dimen@{$\mathchar300\hfil\mathchar301$}\cr
    \noalign{\vskip-.3ex}%
    $#1#2$\cr}}}
\catcode`\@=12 

\newcommand{\IP}{{\rm I$\kern-0.01667em$P}\xspace}

\mathchardef\qsm=63
\mathchardef\pls=43
\mathchardef\mns=512
\mathchardef\plm=518
\mathchardef\eql=61
\mathchardef\smallleft=300
\mathchardef\smallright=301
\mathchardef\les=316
\mathchardef\gre=318
\mathchardef\leq=532
\mathchardef\grq=533

\catcode`\@=11 
\newcounter{pict@width}
\newcounter{pict@height}
\newlength{\pict@scale}
\newcommand{\psfigadd}[4]{%
\setcounter{pict@width}{1*\ratio{#2+\pict@scale/2}{\pict@scale}}
\setcounter{pict@height}{1*\ratio{#3+\pict@scale/2}{\pict@scale}}
\setlength{\unitlength}{\pict@scale}
\hbox to #2{\hspace{-\fill}\begin{picture}(\thepict@width,\thepict@height)
\put(0,0){\psfig{figure=#1,width=#2,height=#3,clip=}}
\SetScale{0.283466457}
\SetWidth{1.763889}
{#4}
\end{picture}}
}
\newcounter{pict@widthfst}
\newcounter{pict@widthscd}
\newcounter{pict@widthtot}
\newcommand{\psfigaddtwo}[7]{%
\setcounter{pict@widthfst}{1*\ratio{#2+\pict@scale/2}{\pict@scale}}
\setcounter{pict@widthscd}{1*\ratio{#2+#4+\pict@scale/2}{\pict@scale}}
\setcounter{pict@widthtot}{1*\ratio{#2+#4+#6+\pict@scale/2}{\pict@scale}}
\setcounter{pict@height}{1*\ratio{#3+\pict@scale/2}{\pict@scale}}
\setlength{\unitlength}{\pict@scale}
\hbox{\hspace{-\fill}\begin{picture}(\thepict@widthtot,\thepict@height)
\put(0,0){\psfig{figure=#1,width=#2,height=#3,clip=}}
\put(\thepict@widthscd,0){\psfig{figure=#5,width=#6,height=#3,clip=}}
\SetScale{0.283466457}
\SetWidth{1.763889}
{#7}
\end{picture}}
}
\newcommand{\psfigror}[4]{%
\setcounter{pict@width}{1*\ratio{#2+\pict@scale/2}{\pict@scale}}
\setcounter{pict@height}{1*\ratio{#3+\pict@scale/2}{\pict@scale}}
\setlength{\unitlength}{\pict@scale}
\hbox{\begin{picture}(\thepict@width,\thepict@height)
\put(0,\thepict@height){\psfig{figure=#1,width=#3,height=#2,clip=,angle=270}}
\SetScale{0.283466457}
\SetWidth{1.763889}
{#4}
\end{picture}}
}
\newcommand{\psfigrol}[4]{%
\setcounter{pict@width}{1*\ratio{#2+\pict@scale/2}{\pict@scale}}
\setcounter{pict@height}{1*\ratio{#3+\pict@scale/2}{\pict@scale}}
\setlength{\unitlength}{\pict@scale}
\hbox{\begin{picture}(\thepict@width,\thepict@height)
\put(0,0){\psfig{figure=#1,width=#3,height=#2,clip=,angle=90}}
\SetScale{0.283466457}
\SetWidth{1.763889}
{#4}
\end{picture}}
}
\catcode`\@=12 
\newlength\listtextwidth

\catcode`\@=11 
\newlength{\@tabfninsert}
\newlength{\@tabfnwidth}
\newcommand{\tabfootnote}[2]{%
  \setlength{\@tabfninsert}{0.8em}
  \setlength{\@tabfnwidth}{\textwidth}
  \addtolength{\@tabfnwidth}{-\@tabfninsert}
  \addtolength{\@tabfnwidth}{-0.4em}
  \noindent\makebox[\@tabfninsert][r]{\footnotesize$^{#1}$\hfil}\hfill%
  \parbox[t]{\@tabfnwidth}{\footnotesize #2\hfill}}
\catcode`\@=12 

\definecolor{darkyellow}{rgb}{0.95,0.95,0.0}

\newcommand{\bec}{\begin{center}}
\newcommand{\bef}{\begin{figure}}
\newcommand{\bet}{\begin{table}}
\newcommand{\bee}{\begin{equation}}
\newcommand{\bei}{\begin{itemize}}
\newcommand{\enc}{\end{center}}
\newcommand{\enf}{\end{figure}}
\newcommand{\ent}{\end{table}}
\newcommand{\ene}{\end{equation}}
\newcommand{\eni}{\end{itemize}}

\def\citeCTD{{\cite{%
nim:a279:290,*npps:b32:181,*nim:a338:254%
}}\xspace}
\def\citeCAL{{\cite{%
nim:a309:77,*nim:a309:101,*nim:a321:356,*nim:a336:23%
}}\xspace}

\begin{document}
\newcommand{\massmu}{M_{\mu \mu}}
\newcommand{\eejet}{{($e$-)$e$-jet}\xspace}
\newcommand{\eemu}{{($e$-)$e$-$\mu$}\xspace}
\newcommand{\emujet}{{($e$-)$\mu$-jet}\xspace}
\newcommand{\ejetjet}{{($e$-)jet-jet}\xspace}
\newcommand{\pom}{I\!\!P}
\prepnum{{DESY--10--250}}
\title{
Study of tau-pair production at HERA
}

\author{ZEUS Collaboration}
\date{December 2010}

\abstract{
A study of events containing two tau leptons with high transverse momentum has
 been performed with the ZEUS detector at HERA, using a data sample
corresponding to an integrated luminosity of 0.33~$\fbi$.
The tau candidates were identified from their decays into electrons, muons or hadronic jets.
The number of tau-pair candidates has been compared with
 the prediction from the Standard Model, where
 the  largest contribution is expected from Bethe-Heitler
 processes. The total visible cross section was extracted.
Standard Model expectations agree well with the measured distributions,
also at high invariant mass of the tau pair.
}

\makezeustitle

\def\3{\ss}

\pagenumbering{Roman}

\begin{center}
{                      \Large  The ZEUS Collaboration              }
\end{center}

%
%
%
%


                                                   %

{\small


{\mbox H.~Abramowicz$^{44, ag}$, }
{\mbox L.~Adamczyk$^{13}$, }
{\mbox M.~Adamus$^{53}$, }
{\mbox R.~Aggarwal$^{7, d}$, }
{\mbox S.~Antonelli$^{4}$, }
{\mbox P.~Antonioli$^{3}$, }
{\mbox A.~Antonov$^{32}$, }
{\mbox M.~Arneodo$^{49}$, }
{\mbox V.~Aushev$^{26}$, }
{\mbox Y.~Aushev$^{26, aa}$
% ...
}
\newpage


\makebox[3em]{$^{1}$}
\begin{minipage}[t]{14cm}
{\it Argonne National Laboratory, Argonne, Illinois 60439-4815, USA}~$^{A}$

\end{minipage}\\
\makebox[3em]{$^{2}$}
\begin{minipage}[t]{14cm}
{\it Andrews University, Berrien Springs, Michigan 49104-0380, USA}

\end{minipage}\\
\makebox[3em]{$^{3}$}
\begin{minipage}[t]{14cm}
{\it INFN Bologna, Bologna, Italy}~$^{B}$

\end{minipage}\\
\makebox[3em]{$^{4}$}
\begin{minipage}[t]{14cm}
{\it University and INFN Bologna, Bologna, Italy}~$^{B}$

\end{minipage}\\
\makebox[3em]{$^{5}$}
\begin{minipage}[t]{14cm}
{\it Physikalisches Institut der Universit\"at Bonn,
Bonn, Germany}~$^{C}$

\end{minipage}\\
\makebox[3em]{$^{6}$}
\begin{minipage}[t]{14cm}
{\it H.H.~Wills Physics Laboratory, University of Bristol,
Bristol, United Kingdom}~$^{D}$

\end{minipage}\\
\makebox[3em]{$^{7}$}
\begin{minipage}[t]{14cm}
{\it Panjab University, Department of Physics, Chandigarh, India}

\end{minipage}\\
\makebox[3em]{$^{8}$}
\begin{minipage}[t]{14cm}
{\it Calabria University,
Physics Department and INFN, Cosenza, Italy}~$^{B}$

\end{minipage}\\
\makebox[3em]{$^{9}$}
\begin{minipage}[t]{14cm}
{\it Institute for Universe and Elementary Particles, Chonnam National University,\\
Kwangju, South Korea}

\end{minipage}\\
\makebox[3em]{$^{10}$}
\begin{minipage}[t]{14cm}
{\it Jabatan Fizik, Universiti Malaya, 50603 Kuala Lumpur, Malaysia}~$^{E}$

\end{minipage}\\
\makebox[3em]{$^{11}$}
\begin{minipage}[t]{14cm}
{\it Nevis Laboratories, Columbia University, Irvington on Hudson,
New York 10027, USA}~$^{F}$

\end{minipage}\\
\makebox[3em]{$^{12}$}
\begin{minipage}[t]{14cm}
{\it The Henryk Niewodniczanski Institute of Nuclear Physics, Polish Academy of \\
Sciences, Cracow, Poland}~$^{G}$

\end{minipage}\\
\makebox[3em]{$^{13}$}
\begin{minipage}[t]{14cm}
{\it Faculty of Physics and Applied Computer Science, AGH-University of Science and \\
Technology, Cracow, Poland}~$^{H}$

\end{minipage}\\
\makebox[3em]{$^{14}$}
\begin{minipage}[t]{14cm}
{\it Department of Physics, Jagellonian University, Cracow, Poland}

\end{minipage}\\
\makebox[3em]{$^{15}$}
\begin{minipage}[t]{14cm}
{\it Deutsches Elektronen-Synchrotron DESY, Hamburg, Germany}

\end{minipage}\\
\makebox[3em]{$^{16}$}
\begin{minipage}[t]{14cm}
{\it Deutsches Elektronen-Synchrotron DESY, Zeuthen, Germany}

\end{minipage}\\
\makebox[3em]{$^{17}$}
\begin{minipage}[t]{14cm}
{\it INFN Florence, Florence, Italy}~$^{B}$

\end{minipage}\\
\makebox[3em]{$^{18}$}
\begin{minipage}[t]{14cm}
{\it University and INFN Florence, Florence, Italy}~$^{B}$

\end{minipage}\\
\makebox[3em]{$^{19}$}
\begin{minipage}[t]{14cm}
{\it Fakult\"at f\"ur Physik der Universit\"at Freiburg i.Br.,
Freiburg i.Br., Germany}

\end{minipage}\\
\makebox[3em]{$^{20}$}
\begin{minipage}[t]{14cm}
{\it School of Physics and Astronomy, University of Glasgow,
Glasgow, United Kingdom}~$^{D}$

\end{minipage}\\
\makebox[3em]{$^{21}$}
\begin{minipage}[t]{14cm}
{\it Department of Engineering in Management and Finance, Univ. of
the Aegean, Chios, Greece}

\end{minipage}\\
\makebox[3em]{$^{22}$}
\begin{minipage}[t]{14cm}
{\it Hamburg University, Institute of Experimental Physics, Hamburg,
Germany}~$^{I}$

\end{minipage}\\
\makebox[3em]{$^{23}$}
\begin{minipage}[t]{14cm}
{\it Imperial College London, High Energy Nuclear Physics Group,
London, United Kingdom}~$^{D}$

\end{minipage}\\
\makebox[3em]{$^{24}$}
\begin{minipage}[t]{14cm}
{\it Institute of Particle and Nuclear Studies, KEK,
Tsukuba, Japan}~$^{J}$

\end{minipage}\\
\makebox[3em]{$^{25}$}
\begin{minipage}[t]{14cm}
{\it Institute of Physics and Technology of Ministry of Education and
Science of Kazakhstan, Almaty, Kazakhstan}

\end{minipage}\\
\makebox[3em]{$^{26}$}
\begin{minipage}[t]{14cm}
{\it Institute for Nuclear Research, National Academy of Sciences, and
National University of Kyiv, Kyiv, Ukraine"}

\end{minipage}\\
\makebox[3em]{$^{27}$}
\begin{minipage}[t]{14cm}
{\it Kyungpook National University, Center for High Energy Physics, Daegu,
South Korea}~$^{K}$

\end{minipage}\\
\makebox[3em]{$^{28}$}
\begin{minipage}[t]{14cm}
{\it Institut de Physique Nucl\'{e}aire, Universit\'{e} Catholique de Louvain, Louvain-la-Neuve,\\
Belgium}~$^{L}$

\end{minipage}\\
\makebox[3em]{$^{29}$}
\begin{minipage}[t]{14cm}
{\it Departamento de F\'{\i}sica Te\'orica, Universidad Aut\'onoma
de Madrid, Madrid, Spain}~$^{M}$

\end{minipage}\\
\makebox[3em]{$^{30}$}
\begin{minipage}[t]{14cm}
{\it Department of Physics, McGill University,
Montr\'eal, Qu\'ebec, Canada H3A 2T8}~$^{N}$

\end{minipage}\\
\makebox[3em]{$^{31}$}
\begin{minipage}[t]{14cm}
{\it Meiji Gakuin University, Faculty of General Education,
Yokohama, Japan}~$^{J}$

\end{minipage}\\
\makebox[3em]{$^{32}$}
\begin{minipage}[t]{14cm}
{\it Moscow Engineering Physics Institute, Moscow, Russia}~$^{O}$

\end{minipage}\\
\makebox[3em]{$^{33}$}
\begin{minipage}[t]{14cm}
{\it Moscow State University, Institute of Nuclear Physics,
Moscow, Russia}~$^{P}$

\end{minipage}\\
\makebox[3em]{$^{34}$}
\begin{minipage}[t]{14cm}
{\it Max-Planck-Institut f\"ur Physik, M\"unchen, Germany}

\end{minipage}\\
\makebox[3em]{$^{35}$}
\begin{minipage}[t]{14cm}
{\it NIKHEF and University of Amsterdam, Amsterdam, Netherlands}~$^{Q}$

\end{minipage}\\
\makebox[3em]{$^{36}$}
\begin{minipage}[t]{14cm}
{\it Physics Department, Ohio State University,
Columbus, Ohio 43210, USA}~$^{A}$

\end{minipage}\\
\makebox[3em]{$^{37}$}
\begin{minipage}[t]{14cm}
{\it Department of Physics, University of Oxford,
Oxford, United Kingdom}~$^{D}$

\end{minipage}\\
\makebox[3em]{$^{38}$}
\begin{minipage}[t]{14cm}
{\it INFN Padova, Padova, Italy}~$^{B}$

\end{minipage}\\
\makebox[3em]{$^{39}$}
\begin{minipage}[t]{14cm}
{\it Dipartimento di Fisica dell' Universit\`a and INFN,
Padova, Italy}~$^{B}$

\end{minipage}\\
\makebox[3em]{$^{40}$}
\begin{minipage}[t]{14cm}
{\it Department of Physics, Pennsylvania State University, University Park,\\
Pennsylvania 16802, USA}~$^{F}$

\end{minipage}\\
\makebox[3em]{$^{41}$}
\begin{minipage}[t]{14cm}
{\it Polytechnic University, Sagamihara, Japan}~$^{J}$

\end{minipage}\\
\makebox[3em]{$^{42}$}
\begin{minipage}[t]{14cm}
{\it Dipartimento di Fisica, Universit\`a 'La Sapienza' and INFN,
Rome, Italy}~$^{B}$

\end{minipage}\\
\makebox[3em]{$^{43}$}
\begin{minipage}[t]{14cm}
{\it Rutherford Appleton Laboratory, Chilton, Didcot, Oxon,
United Kingdom}~$^{D}$

\end{minipage}\\
\makebox[3em]{$^{44}$}
\begin{minipage}[t]{14cm}
{\it Raymond and Beverly Sackler Faculty of Exact Sciences, School of Physics, \\
Tel Aviv University, Tel Aviv, Israel}~$^{R}$

\end{minipage}\\
\makebox[3em]{$^{45}$}
\begin{minipage}[t]{14cm}
{\it Department of Physics, Tokyo Institute of Technology,
Tokyo, Japan}~$^{J}$

\end{minipage}\\
\makebox[3em]{$^{46}$}
\begin{minipage}[t]{14cm}
{\it Department of Physics, University of Tokyo,
Tokyo, Japan}~$^{J}$

\end{minipage}\\
\makebox[3em]{$^{47}$}
\begin{minipage}[t]{14cm}
{\it Tokyo Metropolitan University, Department of Physics,
Tokyo, Japan}~$^{J}$

\end{minipage}\\
\makebox[3em]{$^{48}$}
\begin{minipage}[t]{14cm}
{\it Universit\`a di Torino and INFN, Torino, Italy}~$^{B}$

\end{minipage}\\
\makebox[3em]{$^{49}$}
\begin{minipage}[t]{14cm}
{\it Universit\`a del Piemonte Orientale, Novara, and INFN, Torino,
Italy}~$^{B}$

\end{minipage}\\
\makebox[3em]{$^{50}$}
\begin{minipage}[t]{14cm}
{\it Department of Physics, University of Toronto, Toronto, Ontario,
Canada M5S 1A7}~$^{N}$

\end{minipage}\\
\makebox[3em]{$^{51}$}
\begin{minipage}[t]{14cm}
{\it Physics and Astronomy Department, University College London,
London, United Kingdom}~$^{D}$

\end{minipage}\\
\makebox[3em]{$^{52}$}
\begin{minipage}[t]{14cm}
{\it Faculty of Physics, University of Warsaw, Warsaw, Poland}

\end{minipage}\\
\makebox[3em]{$^{53}$}
\begin{minipage}[t]{14cm}
{\it Institute for Nuclear Studies, Warsaw, Poland}

\end{minipage}\\
\makebox[3em]{$^{54}$}
\begin{minipage}[t]{14cm}
{\it Department of Particle Physics and Astrophysics, Weizmann
Institute, Rehovot, Israel}

\end{minipage}\\
\makebox[3em]{$^{55}$}
\begin{minipage}[t]{14cm}
{\it Department of Physics, University of Wisconsin, Madison,
Wisconsin 53706, USA}~$^{A}$

\end{minipage}\\
\makebox[3em]{$^{56}$}
\begin{minipage}[t]{14cm}
{\it Department of Physics, York University, Ontario, Canada M3J
1P3}~$^{N}$

\end{minipage}\\
\vspace{30em} \pagebreak[4]


\makebox[3ex]{$^{ A}$}
\begin{minipage}[t]{14cm}
 supported by the US Department of Energy\
\end{minipage}\\
\makebox[3ex]{$^{ B}$}
\begin{minipage}[t]{14cm}
 supported by the Italian National Institute for Nuclear Physics (INFN) \
\end{minipage}\\
\makebox[3ex]{$^{ C}$}
\begin{minipage}[t]{14cm}
 supported by the German Federal Ministry for Education and Research (BMBF), under
 contract No. 05 H09PDF\
\end{minipage}\\
\makebox[3ex]{$^{ D}$}
\begin{minipage}[t]{14cm}
 supported by the Science and Technology Facilities Council, UK\
\end{minipage}\\
\makebox[3ex]{$^{ E}$}
\begin{minipage}[t]{14cm}
 supported by an FRGS grant from the Malaysian government\
\end{minipage}\\
\makebox[3ex]{$^{ F}$}
\begin{minipage}[t]{14cm}
 supported by the US National Science Foundation. Any opinion,
 findings and conclusions or recommendations expressed in this material
 are those of the authors and do not necessarily reflect the views of the
 National Science Foundation.\
\end{minipage}\\
\makebox[3ex]{$^{ G}$}
\begin{minipage}[t]{14cm}
 supported by the Polish Ministry of Science and Higher Education as a scientific project No.
 DPN/N188/DESY/2009\
\end{minipage}\\
\makebox[3ex]{$^{ H}$}
\begin{minipage}[t]{14cm}
 supported by the Polish Ministry of Science and Higher Education
 as a scientific project (2009-2010)\
\end{minipage}\\
\makebox[3ex]{$^{ I}$}
\begin{minipage}[t]{14cm}
 supported by the German Federal Ministry for Education and Research (BMBF), under
 contract No. 05h09GUF, and the SFB 676 of the Deutsche Forschungsgemeinschaft (DFG) \
\end{minipage}\\
\makebox[3ex]{$^{ J}$}
\begin{minipage}[t]{14cm}
 supported by the Japanese Ministry of Education, Culture, Sports, Science and Technology
 (MEXT) and its grants for Scientific Research\
\end{minipage}\\
\makebox[3ex]{$^{ K}$}
\begin{minipage}[t]{14cm}
 supported by the Korean Ministry of Education and Korea Science and Engineering
 Foundation\
\end{minipage}\\
\makebox[3ex]{$^{ L}$}
\begin{minipage}[t]{14cm}
 supported by FNRS and its associated funds (IISN and FRIA) and by an Inter-University
 Attraction Poles Programme subsidised by the Belgian Federal Science Policy Office\
\end{minipage}\\
\makebox[3ex]{$^{ M}$}
\begin{minipage}[t]{14cm}
 supported by the Spanish Ministry of Education and Science through funds provided by
 CICYT\
\end{minipage}\\
\makebox[3ex]{$^{ N}$}
\begin{minipage}[t]{14cm}
 supported by the Natural Sciences and Engineering Research Council of Canada (NSERC) \
\end{minipage}\\
\makebox[3ex]{$^{ O}$}
\begin{minipage}[t]{14cm}
 partially supported by the German Federal Ministry for Education and Research (BMBF)\
\end{minipage}\\
\makebox[3ex]{$^{ P}$}
\begin{minipage}[t]{14cm}
 supported by RF Presidential grant N 41-42.2010.2 for the Leading
 Scientific Schools and by the Russian Ministry of Education and Science through its
 grant for Scientific Research on High Energy Physics\
\end{minipage}\\
\makebox[3ex]{$^{ Q}$}
\begin{minipage}[t]{14cm}
 supported by the Netherlands Foundation for Research on Matter (FOM)\
\end{minipage}\\
\makebox[3ex]{$^{ R}$}
\begin{minipage}[t]{14cm}
 supported by the Israel Science Foundation\
\end{minipage}\\
\vspace{30em} \pagebreak[4]


\makebox[3ex]{$^{ a}$}
\begin{minipage}[t]{14cm}
also affiliated with University College London,
 United Kingdom\
\end{minipage}\\
\makebox[3ex]{$^{ b}$}
\begin{minipage}[t]{14cm}
now at University of Salerno, Italy\
\end{minipage}\\
\makebox[3ex]{$^{ c}$}
\begin{minipage}[t]{14cm}
now at Queen Mary University of London, United Kingdom\
\end{minipage}\\
\makebox[3ex]{$^{ d}$}
\begin{minipage}[t]{14cm}
also funded by Max Planck Institute for Physics, Munich, Germany\
\end{minipage}\\
\makebox[3ex]{$^{ e}$}
\begin{minipage}[t]{14cm}
also Senior Alexander von Humboldt Research Fellow at Hamburg University,
 Institute of Experimental Physics, Hamburg, Germany\
\end{minipage}\\
\makebox[3ex]{$^{ f}$}
\begin{minipage}[t]{14cm}
also at Cracow University of Technology, Faculty of Physics,
 Mathemathics and Applied Computer Science, Poland\
\end{minipage}\\
\makebox[3ex]{$^{ g}$}
\begin{minipage}[t]{14cm}
supported by the research grant No. 1 P03B 04529 (2005-2008)\
\end{minipage}\\
\makebox[3ex]{$^{ h}$}
\begin{minipage}[t]{14cm}
now at Rockefeller University, New York, NY
 10065, USA\
\end{minipage}\\
\makebox[3ex]{$^{ i}$}
\begin{minipage}[t]{14cm}
now at DESY group FS-CFEL-1\
\end{minipage}\\
\makebox[3ex]{$^{ j}$}
\begin{minipage}[t]{14cm}
now at Institute of High Energy Physics, Beijing, China\
\end{minipage}\\
\makebox[3ex]{$^{ k}$}
\begin{minipage}[t]{14cm}
now at DESY group FEB, Hamburg, Germany\
\end{minipage}\\
\makebox[3ex]{$^{ l}$}
\begin{minipage}[t]{14cm}
also at Moscow State University, Russia\
\end{minipage}\\
\makebox[3ex]{$^{ m}$}
\begin{minipage}[t]{14cm}
now at University of Liverpool, United Kingdom\
\end{minipage}\\
\makebox[3ex]{$^{ n}$}
\begin{minipage}[t]{14cm}
now at CERN, Geneva, Switzerland\
\end{minipage}\\
\makebox[3ex]{$^{ o}$}
\begin{minipage}[t]{14cm}
also affiliated with Universtiy College London, UK\
\end{minipage}\\
\makebox[3ex]{$^{ p}$}
\begin{minipage}[t]{14cm}
now at Goldman Sachs, London, UK\
\end{minipage}\\
\makebox[3ex]{$^{ q}$}
\begin{minipage}[t]{14cm}
also at Institute of Theoretical and Experimental Physics, Moscow, Russia\
\end{minipage}\\
\makebox[3ex]{$^{ r}$}
\begin{minipage}[t]{14cm}
also at INP, Cracow, Poland\
\end{minipage}\\
\makebox[3ex]{$^{ s}$}
\begin{minipage}[t]{14cm}
also at FPACS, AGH-UST, Cracow, Poland\
\end{minipage}\\
\makebox[3ex]{$^{ t}$}
\begin{minipage}[t]{14cm}
partially supported by Warsaw University, Poland\
\end{minipage}\\
\makebox[3ex]{$^{ u}$}
\begin{minipage}[t]{14cm}
now at Istituto Nucleare di Fisica Nazionale (INFN), Pisa, Italy\
\end{minipage}\\
\makebox[3ex]{$^{ v}$}
\begin{minipage}[t]{14cm}
now at Haase Energie Technik AG, Neum\"unster, Germany\
\end{minipage}\\
\makebox[3ex]{$^{ w}$}
\begin{minipage}[t]{14cm}
now at Department of Physics, University of Bonn, Germany\
\end{minipage}\\
\makebox[3ex]{$^{ x}$}
\begin{minipage}[t]{14cm}
also affiliated with DESY, Germany\
\end{minipage}\\
\makebox[3ex]{$^{ y}$}
\begin{minipage}[t]{14cm}
also at University of Tokyo, Japan\
\end{minipage}\\
\makebox[3ex]{$^{ z}$}
\begin{minipage}[t]{14cm}
now at Kobe University, Japan\
\end{minipage}\\
\makebox[3ex]{$^{\dagger}$}
\begin{minipage}[t]{14cm}
 deceased \
\end{minipage}\\
\makebox[3ex]{$^{aa}$}
\begin{minipage}[t]{14cm}
supported by DESY, Germany\
\end{minipage}\\
\makebox[3ex]{$^{ab}$}
\begin{minipage}[t]{14cm}
member of National University of Kyiv - Mohyla Academy, Kyiv, Ukraine\
\end{minipage}\\
\makebox[3ex]{$^{ac}$}
\begin{minipage}[t]{14cm}
supported by the Bogolyubov Institute for Theoretical Physics of the National
 Academy of Sciences, Ukraine\
\end{minipage}\\
\makebox[3ex]{$^{ad}$}
\begin{minipage}[t]{14cm}
STFC Advanced Fellow\
\end{minipage}\\
\makebox[3ex]{$^{ae}$}
\begin{minipage}[t]{14cm}
nee Korcsak-Gorzo\
\end{minipage}\\
\makebox[3ex]{$^{af}$}
\begin{minipage}[t]{14cm}
This material was based on work supported by the
 National Science Foundation, while working at the Foundation.\
\end{minipage}\\
\makebox[3ex]{$^{ag}$}
\begin{minipage}[t]{14cm}
also at Max Planck Institute for Physics, Munich, Germany, External Scientific Member\
\end{minipage}\\
\makebox[3ex]{$^{ah}$}
\begin{minipage}[t]{14cm}
now at Tokyo Metropolitan University, Japan\
\end{minipage}\\
\makebox[3ex]{$^{ai}$}
\begin{minipage}[t]{14cm}
now at Nihon Institute of Medical Science, Japan\
\end{minipage}\\
\makebox[3ex]{$^{aj}$}
\begin{minipage}[t]{14cm}
now at Osaka University, Osaka, Japan\
\end{minipage}\\
\makebox[3ex]{$^{ak}$}
\begin{minipage}[t]{14cm}
also at \L\'{o}d\'{z} University, Poland\
\end{minipage}\\
\makebox[3ex]{$^{al}$}
\begin{minipage}[t]{14cm}
member of \L\'{o}d\'{z} University, Poland\
\end{minipage}\\
\makebox[3ex]{$^{am}$}
\begin{minipage}[t]{14cm}
now at Lund University, Lund, Sweden\
\end{minipage}\\
\makebox[3ex]{$^{an}$}
\begin{minipage}[t]{14cm}
also at University of Podlasie, Siedlce, Poland\
\end{minipage}\\

}


\newpage

\pagenumbering{arabic} 
\pagestyle{plain}
\section{Introduction}
\label{sec-int}

The production
of isolated-lepton pairs at the $ep$ collider HERA is dominated by the two-photon Bethe-Heitler process,
$\gamma \gamma \to  l^+ l^-$, and 
can be accurately predicted in the Standard Model (SM)~\cite{zfp:c52:289}. 
Possible deviations of the event yield or final-state distributions from
the prediction of the SM could be a hint for new physics.
The measurement of multi-lepton production at HERA attracted some
interest, especially after the observation of an excess of events at high mass
in multi-electron final states, observed by the H1 Collaboration in the
HERA I (1994--2000) data~\cite{epj:c31:17}. 
Recently, the H1 and ZEUS  Collaborations have published~\cite{Desy-08-065,Desy-09-072,Desy-09-108} a study of multi-electron and multi-muon events based on the entire statistics
collected at HERA.
Further investigations of multi-lepton events were performed  
in tau-pair production by H1~\cite{epj:c48:699}, based on the HERA I data.

This paper reports a study of ditau events with the ZEUS detector, with
data collected in the HERA II running phase (2004--2007).
The taus are identified from their decay into an
electron, a muon or a hadronic jet.
The hadronic
channel is selected with a technique analogous to that used in a
previous ZEUS publication on single-tau production~\cite{pl:b583:41}, where two interesting events, with
a high-transverse-energy tau candidate and  large missing transverse momentum, 
were observed in the HERA I data.

\section{Experimental set-up}
\label{sec-expandsim}

 The data were collected between 2004 and 2007 at the $ep$
 collider HERA using the ZEUS detector. During this period HERA operated 
 with an electron or positron\footnote{Here and in the following, the term
 ``electron'' denotes generically both the electron ($e^-$) and the 
 positron ($e^+$), unless otherwise specified.} beam with an  energy of 27.5\,\gev~ and 
a proton beam  with an
 energy of 920\,\gev. The $e^-p$ data correspond to an integrated luminosity of 179\,$\pbi$, 
while the $e^+p$ collisions correspond to 155\,$\pbi$, giving a total of 334\,$\pbi$. The lepton beams were polarised, with roughly equal periods
for right-handed and left-handed polarisation, such that the average polarisation was negligible.

\Zdetdesc

 Charged particles were tracked in the central tracking detector (CTD)~\citeCTD,
 which operated in a magnetic field of $1.43\Tesla$ provided by a thin
 superconducting solenoid. The CTD consisted of 72~cylindrical drift chamber
 layers, organised in nine superlayers covering the
 polar-angle\footnote{The ZEUS coordinate system is a right-handed Cartesian
 system, with the $Z$ axis pointing in the proton beam direction,
 referred to as the ``forward direction'',
 and the $X$ axis pointing left towards the centre of HERA. The polar angle, $\theta$, is
measured with respect to the proton beam direction. The coordinate origin is at
the nominal interaction point.} 
region \mbox{$15^\circ<\theta<164^\circ$}. 
The CTD was complemented by a silicon microvertex detector
 (MVD) \cite{nim:a581:656}, consisting of three active layers in the barrel
and four disks in the forward region.

The high-resolution uranium--scintillator calorimeter (CAL)~\citeCAL consisted 
of three parts: the forward (FCAL), the barrel (BCAL) and the rear (RCAL)
calorimeters. Each part was subdivided transversely into towers and
longitudinally into one electromagnetic section (EMC) and either one (in RCAL)
or two (in BCAL and FCAL) hadronic sections (HAC). The smallest subdivision of
the calorimeter was called a cell.  The CAL energy resolutions, as measured under
test-beam conditions, were $\sigma(E)/E=0.18/\sqrt{E}$ for electrons and
$\sigma(E)/E=0.35/\sqrt{E}$ for hadrons, with $E$ in $\Gev$.

The muon system consisted of barrel, rear (B/RMUON) \cite{zeus:1993:bluebook} and forward (FMUON)
tracking detectors. The BMUON (RMUON) consisted of limited-streamer (LS) tube
chambers placed behind the BCAL (RCAL), inside and outside a magnetised iron
yoke surrounding the CAL, covering the polar-angle region 
$34^\circ < \theta < 135^\circ$ ($135^\circ < \theta < 171^\circ$). The FMUON
consisted of six trigger planes of LS tubes and four planes of drift
chambers covering the polar-angle region $5^\circ < \theta < 32^\circ$.

 The luminosity was measured using the Bethe-Heitler reaction 
$ep  \to e \gamma p$ by a luminosity detector which consisted of 
a lead--scintillator calorimeter~\cite{Desy-92-066,*zfp:c63:391,*acpp:b32:2025}
and an independent magnetic spectrometer~\cite{nim:a565:572}.
The fractional uncertainty on the measured luminosity was $1.9\%$.

\section{Signal and background processes}

\label{sec-simulation}

The signal considered in this analysis is the presence of two
isolated taus at high transverse energy from the reaction
$e p \to (e) \tau^+ \tau^- X$. 
The scattered electron is observed in the calorimeter only for
high virtuality of the photon at the electron vertex, $Q^2$.   
The proton either stays intact (elastic reaction, $X=p$) or 
dissociates into a resonant (quasi-elastic) or hadronic system (inelastic).
In order to suppress the dominant backgrounds from 
deep inelastic scattering (DIS) neutral current (NC), $ep \to eX$, and photoproduction, $\gamma p \to X$, 
only events where no deposit was observed in the forward part of the calorimeter were selected.
The dominant signal process 
($\simeq 71\%$ estimated from Monte Carlo simulation after all selection cuts) was therefore the
elastic  reaction $ep \to (e)p \tau^+ \tau^-$, where the final-state proton stays intact.
The quasi-elastic reaction was the second most important contribution ($\simeq 27\%$). 
In this case the final-state proton dissociates into a system with small invariant
mass, escaping
in the forward beam pipe.
The contribution of events $ep \to e \tau^+ \tau^- X$, in which the scattered electron was observed in the
calorimeter, was determined to be only $\simeq 1\%$ after all analysis cuts.

The two tau leptons were identified from their decays into an electron, a muon
or hadrons, respectively, resulting in  the final-state signatures 
listed in Table~\ref{tab-cuts}. The topologies in which the two tau leptons both decayed into
either electrons or muons were not considered, due to the irreducible background of
dielectron and dimuon processes. Hadronic decays give rise to narrow and
low-multiplicity jets; these characteristics allowed 
tau decays to be distinguished from the much more abundant QCD-induced 
jets, as described in Section~\ref{tauidentification}.

Monte Carlo (MC) programs were used to generate the
signal and background events in order to optimise the selection cuts
and determine acceptances.

The {\sc Grape}  event generator \cite{cpc:136:126} was used
to simulate signal events.
It is based on the exact electroweak matrix elements in photon-photon 
(and also photon-$Z^0$ and $Z^0$-$Z^0$) collisions and 
internal photon or $Z^0$ conversions at tree level.
The three contributions at the proton vertex, elastic, quasi-elastic and
inelastic, were generated separately. The cross section 
for tau-pair production
falls steeply with the transverse energy of the tau lepton.

Dielectron events $ep \to (e) e^+ e^-  X$  were a potential background
especially to the topology \eejet and were
also generated with the {\sc Grape} program. Dimuon events, which represented the main
background to the \eemu and \emujet final states, 
were also generated with {\sc Grape}.

Due to the requirement that the events be (quasi-)elastic, the main background 
to the hadronic channel consisted
of diffractive dijet production. 
Diffractive DIS  production, which was the main background to
the topologies \ejetjet and
\eejet, was generated with the {\sc Rapgap}~\cite{cpc:86:147} program. The same program was used to generate
diffractive dijet events in the photoproduction regime, both in 
resolved and direct photon processes, which represented the main background to the 
jet-jet topology. Since  the MC events did not adequately
describe the data distributions,  
the resolved and direct contributions were separately normalised to the data using an
independent sample from that used in the analysis (see Section~\ref{finalselection}). 
Non-diffractive DIS and photoproduction events were generated with 
the 
{\sc Djangoh}~\cite{spi:www:djangoh11} and {\sc Pythia}~\cite{cpc:135:238} programs, respectively.

The generated events were passed through the {\sc Geant} 
3.21-based~\cite{tech:cern-dd-ee-84-1} ZEUS detector- and 
trigger-simulation programs. They were reconstructed and
analysed using the same program chain as the data.

\section{Event selection}
\label{sec-selection}

The events were selected online by  
the ZEUS three-level trigger system  \cite{zeus:1993:bluebook,nim:a580:1257},
using a combination of several trigger chains which required typically  
either the presence of hadronic jets, an electron or a muon in the final state. 
The trigger requirements were looser than the offline selection.
The  offline selection proceeded in two steps~\cite{maeda:phd:2009}. A preselection required 
low track multiplicity and no energy around the beam-pipe hole in the forward
region of the detector, 
as expected for (quasi-)elastic ditau production. 
A second step required the presence of at least two objects among electrons, muons
or jets, identified as the tau decays,  and classified the events in the categories
listed in Table~\ref{tab-cuts}.
This selection is described in more detail below.

\subsection{Preselection}
\label{ssec-preselection}

The following offline criteria were imposed at preselection level:
\begin{itemize}
\item the number of good tracks in the event, $N_\mathrm{trk}$, was required to
be at least 2 and at most 7, as expected for the ditau topologies studied. A good track
was defined to pass through at least 3 CTD superlayers, to have hits in the MVD 
or in the innermost CTD
superlayer and to have a transverse momentum greater than 150~\mev;
\item the $Z$ coordinate of the interaction vertex, reconstructed using tracks, 
was restricted to  $|Z_\mathrm{VTX}|<40~\mathrm{cm}$ in order to reject the background due to non-$ep$ interactions; 
\item the energy $E_\mathrm{FCAL}^\mathrm{IR}$, reconstructed from the sum of the energy 
deposits in the CAL cells in the first inner ring around the forward beam-pipe hole, was restricted to be less than 1 GeV in order to select (quasi-)elastic events;
\item the $E-P_Z$ of the final state, reconstructed from the sum of the total 
and longitudinal energy deposits of the cells in the calorimeter, was required to be
less than 60 GeV. For events with a muon, the $(E-P_Z)$ of the CAL deposit associated
with the muon was replaced by that of the muon track.
For fully contained events, $E-P_Z$ is twice the electron-beam energy and peaks at 55 GeV. This requirement rejected $ep$ interactions overlapping with background events.

\end{itemize}

Other selection criteria were applied to reject residual non-$ep$ interactions, 
mainly beam-gas events and cosmic rays.
It was verified that the loss of signal events due to these cuts was negligible.

\subsection{Identification of electrons and muons}



An algorithm which combined information from the energy deposits in the
calorimeter with tracks~\cite{zfp:c74:207} was used to identify possible electron 
candidates. The electron four-momenta were reconstructed from the CAL.
The electron candidates   were required to have a transverse energy
$p_T^e>2~\gev$, to be in the polar-angle range 
$17^\circ < \theta_e < 160^\circ$ and 
to have a good track matched to
the calorimeter deposit. 
The matched track was required to have an extrapolated  
distance of closest approach to the calorimeter deposit of less than 8 cm.
 The electron candidate was required to be isolated such that the total
energy not associated with the electron in an $\eta$-$\phi$ cone of radius
0.8 centred on the electron direction,  where $\phi$ is the
azimuthal angle and $\eta$ is the pseudorapidity, 
 was less than 2~\gev. This requirement was
complemented by the requirement that no track, other than the matching track,
was contained in an $\eta$-$\phi$ cone of radius 1 centred on the
electron direction. 
Further fiducial cuts~\cite{maeda:phd:2009} were applied in the RCAL
to guarantee that the experimental acceptance was well understood. The charge of the track matched to the electron, $Q_e$,
 was also used to discriminate the signal from the background. The track charge information was used
only if its  significance
$S^\mathrm{trk}=|Q/r|/\sigma(Q/r)$ was greater than 1.5, where $r$ denotes the radius of the track helix and $\sigma$ is the uncertainty.

The muons were required to be reconstructed in the rear or barrel 
muon chambers and to be matched to a good track and 
to a calorimeter deposit.
The muon transverse momentum and direction were reconstructed from the matched track.
Each muon candidate was required to have a transverse momentum 
$p_T^\mu>2~\mathrm{GeV}$ and
 to lie in the angular region 
$34^\circ < \theta_\mu < 157^\circ$. The muon was required to
be isolated such that only the matching track was contained in an $\eta$-$\phi$ cone of radius 1 centred on the muon direction. 
If a second muon candidate, reconstructed with looser criteria, was found in the event,
this event was rejected.

\subsection{Identification of hadronic tau decays}
\label{tauidentification}

The jets deriving from the hadronic tau decay were reconstructed 
from the CAL cells using the $k_T$ cluster algorithm~\cite{np:b406:187} 
in the longitudinally  invariant inclusive mode~\cite{pr:d48:3160}, assuming
massless objects, and were corrected for
energy loss due to the dead material in front of the CAL.  
The jets were required to have a transverse energy $E_T^\mathrm{jet}>5~\gev$  and
 pseudorapidity $|\eta^\mathrm{jet}|<2$. 
At least one good track associated with the jet was required in
an $\eta$-$\phi$ cone of radius 1 around the jet axis.
The fraction $f_{\mathrm{EMC}}$ of the jet energy in the electromagnetic section of the calorimeter was required to satisfy $f_\mathrm{EMC}<\min(0.95,2\cdot R_{90\%}+0.7)$, where $R_{90\%}$ is the radius of
the $\eta$-$\phi$ cone centred on the jet axis that contains $90\%$ of the
jet energy. These two cuts rejected electrons faking hadronic tau decays. 
Further fiducial cuts~\cite{maeda:phd:2009}
were applied on the jet direction in order to 
exclude regions of the CAL where the jet energy was not precisely measured.

Jets originating from hadronic tau decays are characterised by 
their low mass, low multiplicity (mostly either one or three tracks) and 
pencil-like shape. 
In contrast, QCD-induced hadronic jets are typically broader and have higher
multiplicity.
These properties were exploited to discriminate tau jets from 
QCD jets using a multi-variate discrimination 
technique~\cite{nim:a501:576} which was used in
a previous publication~\cite{pl:b583:41}. The jet shape
was characterised by six variables. Five of these variables were the
same as those used in previous analyses~\cite{pl:b583:41,epj:c44:463}: 
the first and the second moment of the radial 
extension of the jet-energy deposition ($R_\mathrm{mean}$ and $R_\mathrm{rms}$,
respectively); the
first moment of the energy 
deposition in the direction along the jet
axis ($L_\mathrm{mean}$); the number of subjets  within the jet resolved 
with a resolution criterion $y_\mathrm{cut}$~\cite{jhep:09:009} 
of $5 \cdot 10^{-4}$ ($N_\mathrm{subj}$); 
and the mass  of the jet calculated from the CAL cells associated with the jet
($M_\mathrm{jet}$).
The sixth variable, which was used for this analysis,
was 
 the sum of the distances in the $\eta$-$\phi$ plane between the jet axis and the tracks associated with the jet, ${R}_\mathrm{trk}=
\sum_i^\mathrm{N_{trk}} \sqrt{(\Delta \eta_i^2+\Delta \phi_i^2)}$.

The
six variables were combined in a discriminant $\mathcal{D}$ given, for each point in the
phase space $\vec{x}(-\log(R_\mathrm{mean}),-\log(R_\mathrm{rms})$,$-\log(1-L_\mathrm{mean}),N_\mathrm{subj},M_\mathrm{jet},\log({R}_\mathrm{trk}))$, by:
$$ \mathcal{D}(\vec{x})= \frac{\rho_\mathrm{sig}(\vec{x})}
                     {\rho_\mathrm{sig}(\vec{x})+\rho_\mathrm{bkg}(\vec{x})},
$$                     
where $\rho_\mathrm{sig}$ and $\rho_\mathrm{bkg}$ are the density functions of the signal and the background, respectively. The densities were calculated from a method based on range searching~\cite{nim:a501:576} and were determined from a sample of single-tau MC events for the signal and  {\sc Djangoh} DIS NC events for the
background. 
For any given jet with phase-space coordinate $\vec{x}$, the signal
and the background densities were evaluated from the number of
corresponding simulated signal and background jets in a 6-dimensional box of
fixed size centred around $\vec{x}$.

\begin{figure}
\begin{center}
\includegraphics[height=18cm]{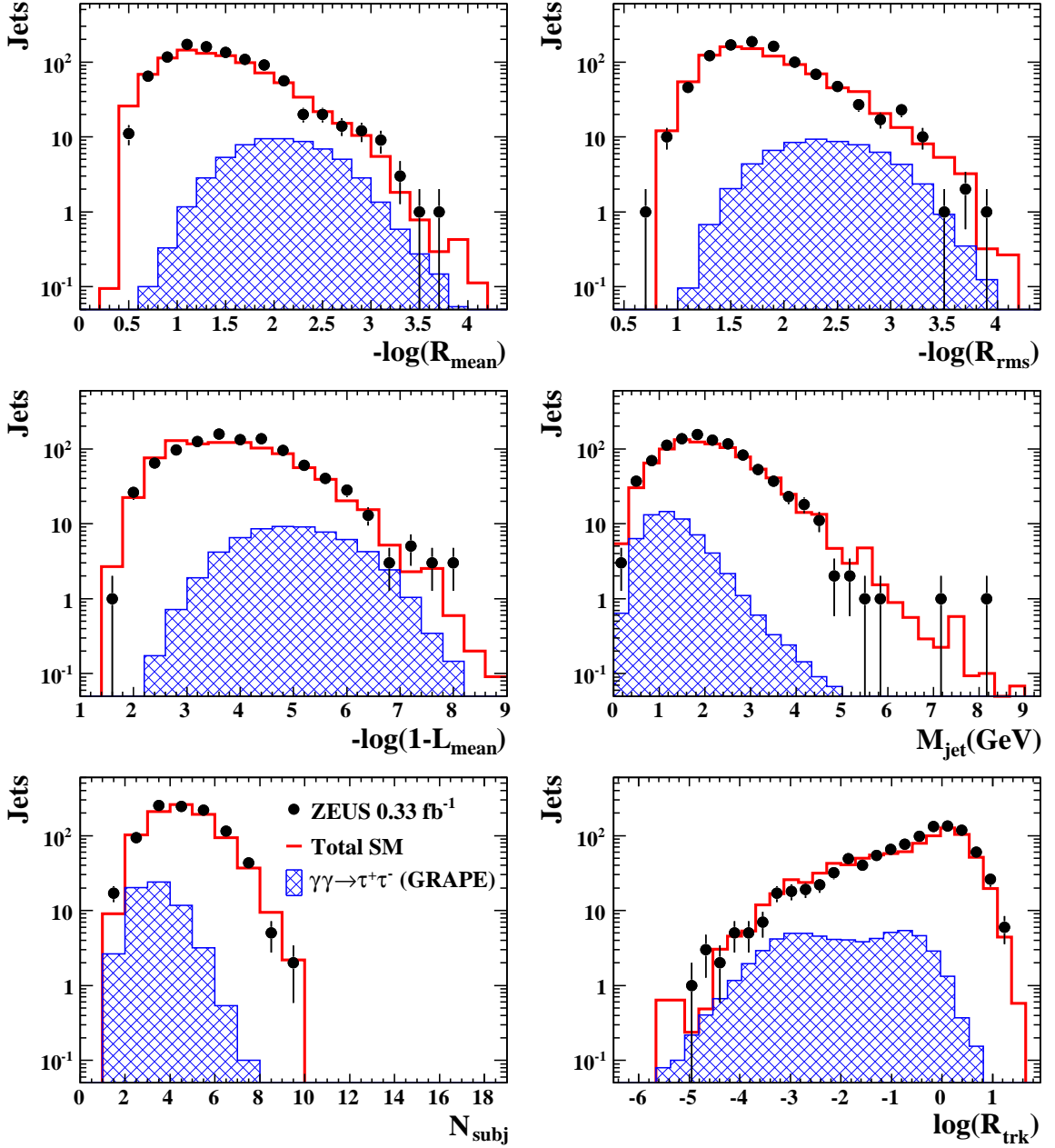}
\end{center}
\caption{
Distributions of the variables used for the tau-jet discriminant,
 for the data (dots) and the sum of
the MC expectations (solid line), after all selection criteria except for the
discriminant cut. The variables are defined in the text.
The data are shown with the statistical uncertainty (vertical
error bars). 
The contribution of the ditau signal as predicted by  {\sc Grape} is shown
separately (hatched histogram). 
The background due to photoproduction interactions is normalised with the
procedure described in the text.
The other MC
expectations are normalised to the luminosity of the data. 
}
\label{fig-discrvariables}
\end{figure}

\begin{figure}
\begin{center}
\includegraphics[height=14cm]{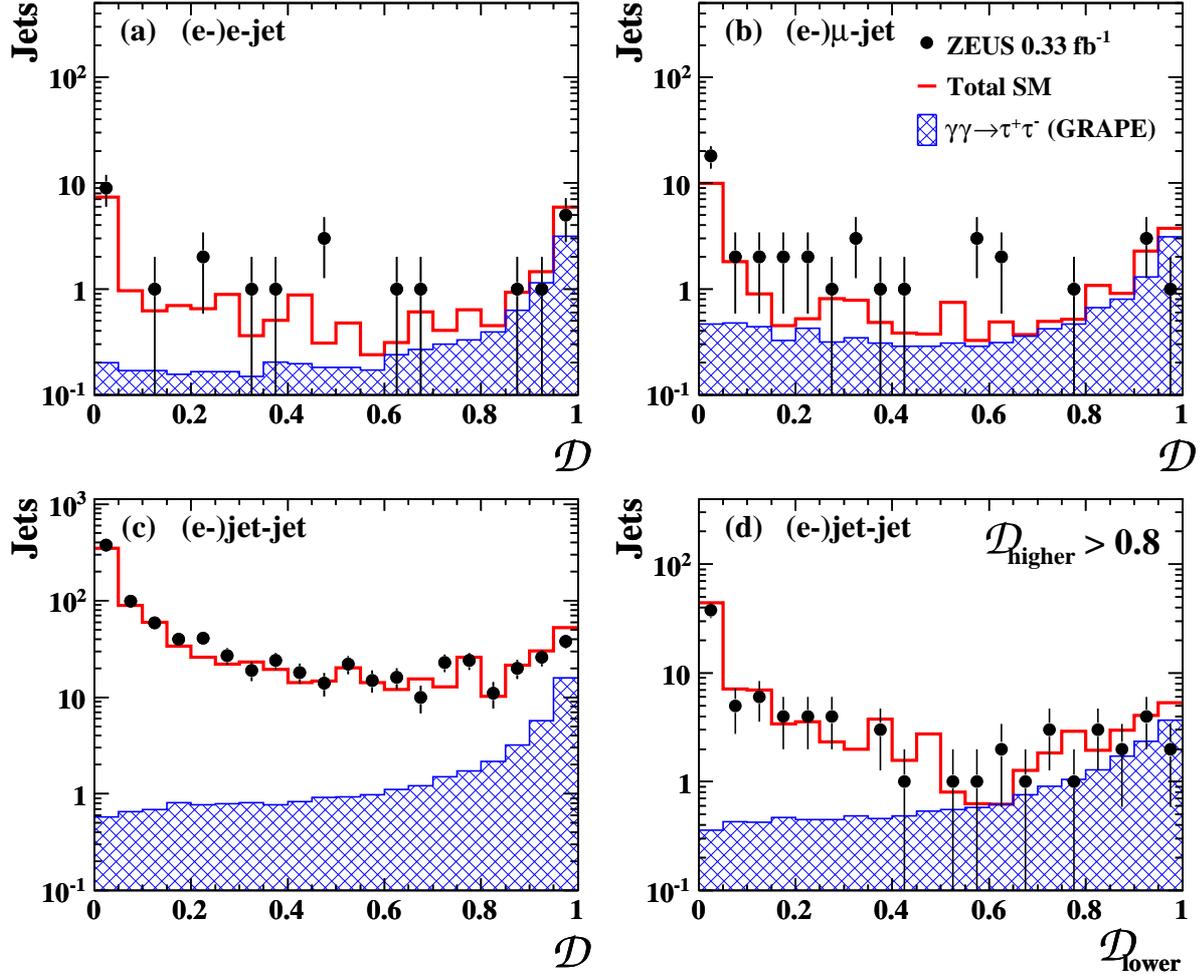}
\end{center}
\caption{Tau-jet
discriminant values for the three decay topologies
(a) \eejet, (b) \emujet and (c) \ejetjet for the events after
all selection criteria, except the discriminant cut. For the \ejetjet channel,
(d) shows the discriminant distribution for the jet with lower $\cal D$ value,
after the requirement ${\cal D}>0.8$ for the jet with higher $\cal D$.
Other details as in the caption to Fig. 1. }
\label{fig-discriminant}
\end{figure}

Figure~\ref{fig-discrvariables} shows the six discriminant variables
 for the data, compared to those of the sum of the MC expectations,
where all cuts described in this Section, except the
discriminant cut, were applied.
 The MC agrees well with the data, both in shape and in normalisation.
The discriminant variable  is shown in Fig.~\ref{fig-discriminant} for each jet 
in each decay channel. As expected,
the ditau signal MC dominates at large values of $\mathcal{D}$, while the background
from the other processes populates the lower values of $\mathcal{D}$.
In order to select the hadronic tau decays, a cut on the discriminant  greater than 0.8 
was applied on each tau-candidate jet. 

\subsection{Final selection}
\label{finalselection}

After the preselection cuts, the possible decay products of each tau
were searched for and
the final state was classified in the eight exclusive topologies listed
in Table~\ref{tab-cuts}, in which each tau of the pair could decay
into an electron, a muon, or a hadronic jet. 
For high $Q^2$, 
the scattered electron could also be observed in the CAL, giving an
additional electron in the topology. For lower $Q^2$, the scattered
electron escaped down the beam pipe in the electron-beam direction. In
this case, the quantity $E-P_Z$ is typically much less than 55 GeV and a cut,
$E-P_Z<45\,\gev$, was applied to reduce the DIS NC background.

The following additional cuts, also listed in Table~\ref{tab-cuts}, were applied
in each case:

\begin{itemize}
\item in the \eemu topology,  exactly one muon, (two) one electron(s) and no
additional tracks were required
in the final state. In the $e$-$\mu$ topology, 
the electron charge, $Q_e$, was required to be opposite
to that of the muon track, $Q_\mu$.
In addition,
in order to reduce the main background due to dimuon production at high
$Q^2$, where one of the muons was outside the acceptance of the 
detector, 
the electron was required to have charge opposite
to that of the electron beam, $Q_\mathrm{beam}$, if $\theta_e>1.0$ and $S^\mathrm{trk}>1.5$;

\item in the \eejet channel, (two) one electron(s) were required
in the event together with exactly one tau-candidate jet. 
In the $e$-jet channel, 
the cuts on the electron were the same as in the $e$-$\mu$ channel, in order to reject the
dominant background  due to DIS NC events $ep \to eX$.
In addition, the charge 
$Q_{\mathrm{jet}}=\sum_i Q_{i,\mathrm{trk}}$ was
reconstructed for the jet, summing the charges $Q_{i,\mathrm{trk}}$ of  all tracks, with significance $S^\mathrm{trk}>1.5$,
associated with the jet. If all tracks satisfied the $S^\mathrm{trk}$ criterion, 
the jet charge was required to be $Q_\mathrm{jet}=\pm 1$ and to be opposite to that of the
electron candidate, as expected in the production of a tau pair;

\item in the \emujet topology, exactly (one electron) one muon
and one tau-candidate jet were required to be present;

\item 
for the  \ejetjet topology, exactly two candidate jets, and in the high-$Q^2$ topology an
additional electron, were required.
The jet-jet channel was dominated by the diffractive photoproduction dijet background and, in order
to suppress it,
the charge $Q_\mathrm{jet}$ was required to be
$\pm 1$ with the two jets having opposite charges.
The $S^\mathrm{trk}$ requirement was the same as in the $e$-jet channel.

\end{itemize}

\begin{figure}
\begin{center}
\includegraphics[height=14cm]{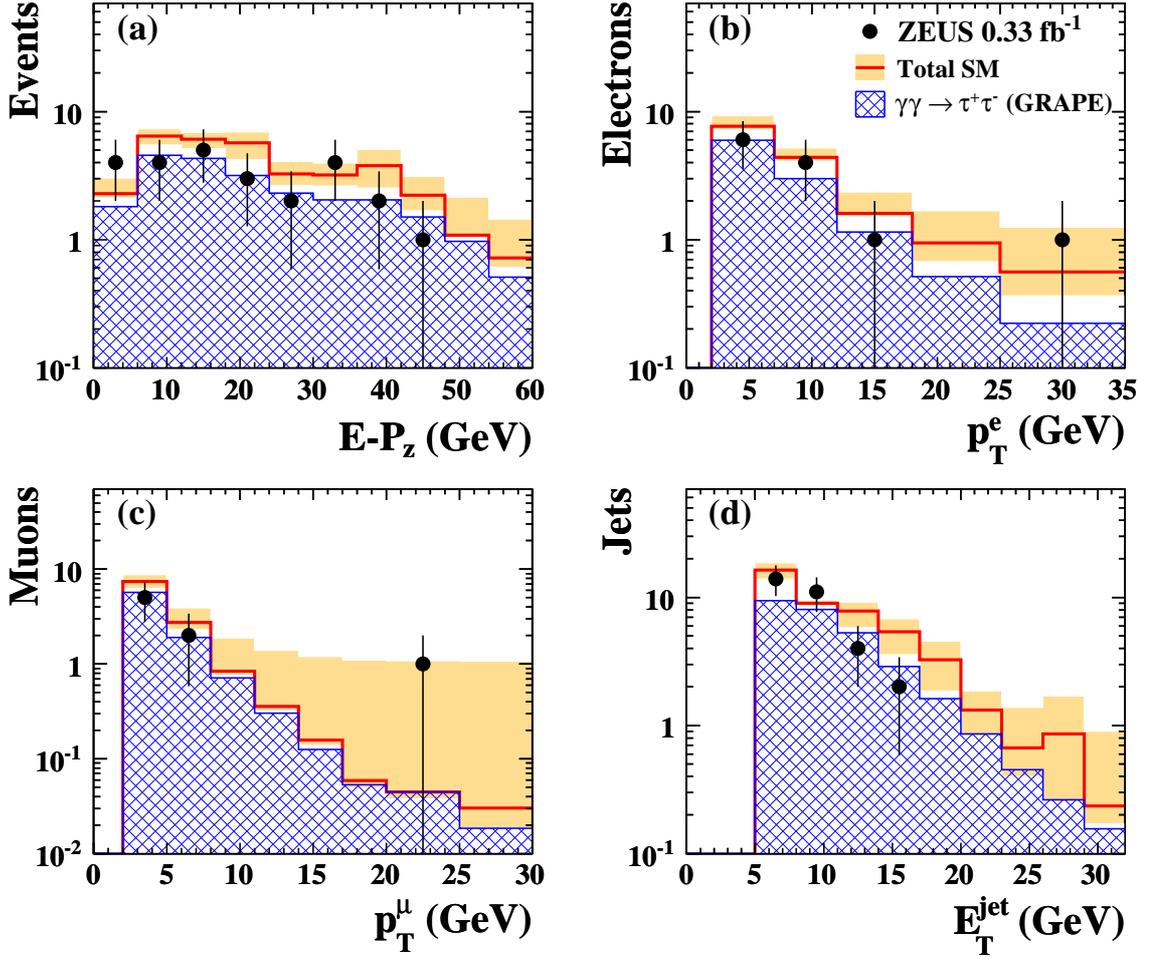}
\end{center}
\caption{Distributions of the selected events  
compared with the predictions
 from the Standard Model (SM). 
 The plots show: (a) the variable $E-P_Z$, 
(b) the transverse momentum, $p_T^\mathrm{e}$, of the electrons,  
(c) the transverse momentum, $p_T^{\mu}$, of the muons 
and (d) the transverse energy of each jet, $E_T^\mathrm{jet}$.
The shaded bands show the systematic 
uncertainty on the SM expectation.
Other details as in the caption to Fig. 1. 
}
\label{fig-distributions}
\end{figure}

In total, 25 events were selected.
Figure~\ref{fig-distributions} shows the $E-P_Z$ distribution for the events  and  the
distributions of  the
transverse momentum of the electron, muon and jet candidates, compared to the sum 
of the signal and background
expectations.

All MC background expectations were normalised to the data luminosity, except
the resolved and direct photoproduction contributions, both diffractive and non-diffractive.
These four contributions were fitted to the
data using two variables~\cite{maeda:phd:2009}, which helped to distinguish the different
processes:
$x_{\gamma}^\mathrm{obs}$, which is an estimator of
the fraction of photon's momentum taking part in the hard interaction;
and $x_{\pom}^\mathrm{obs}$, an estimator of
the fraction of longitudinal momentum transferred from the proton
in diffractive events.
The four normalisation factors were determined in two steps and
for this purpose
an independent data sample was used, 
which required the presence of two jets, $E-P_Z<40~\mathrm{GeV}$ and $8 \leq N_\mathrm{trk} \leq 17$. 
The relative normalisation of the diffractive and non-diffractive
MC contributions was determined first from a fit to the data of the $x_{\pom}^\mathrm{obs}$ 
distribution for $x_{\gamma}^\mathrm{obs}<0.5$. The four MC components were then
fitted to the data $x_{\gamma}^\mathrm{obs}$ variable.
The resulting normalisation factors were around 2, with an uncertainty
of $25\%$ and $15\%$ for the diffractive and non-diffractive processes, respectively.

As shown in Fig.~\ref{fig-distributions}, the MC gives a good
description of the data, both in shape and normalisation.
The diffractive DIS dijet process is the main background in the $e$-jet-jet channel,
while the diffractive photoproduction dijet process is the main background in
the jet-jet channel. The dimuon production process is the main background to
the topologies with a muon in the final state. In the $e$-jet topology, both
the dielectron and the DIS diffractive processes contribute to the background.

\subsection{Systematic uncertainties}
\label{ssec-syst}

The following sources of systematic uncertainties were considered 
(the resulting uncertainty on the total cross section is given in parentheses):
\begin{itemize}
  \item the electron energy scale was changed by its uncertainty of $2\%$  
($^{+0.2}_{-1.5} \%$);

  \item the hadronic jet energy scale was varied by its uncertainty of $3\%$ ($^{+26}_{-7} \%$);

  \item 
the normalisation of the direct and resolved photoproduction background contributions, 
diffractive and
non-diffractive, gave one of the
main contributions to the uncertainty of the MC background expectation.
The diffractive and non-diffractive normalisation factors were changed 
by their uncertainty of
$25\%$ and $15\%$, respectively ($\pm 11\%$);

\item the total muon acceptance, including the trigger, the
reconstruction and the muon identification efficiencies, is known to about
$7\%$~\cite{Desy-09-072} ($\pm 4\%$);

\item the cut on the track charge significance, $S^\mathrm{trk}$, was varied by $\pm 0.5$ 
($^{-12}_{-7}\%$);

\item the cut on the energy in the inner ring of the FCAL was varied ($^{+6}_{-12}\%$);

\item in the $(e)$-$\mu$-jet channel, 
in order to account for the observed discrepancy in the discriminant distribution (see 
Fig.~\ref{fig-discriminant}b), 
the Bethe-Heitler background was increased by $80\%$ ($-6\%$);

\item the effect of potential differences between data and MC in the single-track
efficiency~\cite{ramoona:phd:2011}, 
{\em e.g.}, from secondary interactions in the detector material, was evaluated
($+5\%$);

\item the overall normalisation uncertainty associated with the luminosity
measurement was added ($\pm 1.9\%$); 

\item the statistical uncertainty of the few MC events that survive the cuts was added to the statistical uncertainty in the cross section.
\end{itemize}

A further check was performed on the tau-jet discriminant value. 
The $\rho_\mathrm{bkg}$ density used to determine
$\cal{D}$ was calculated using as background the following different MC samples: 
the diffractive
DIS dijet MC, the diffractive photoproduction dijet sample 
and the inclusive DIS NC 
Monte Carlo events. 
All samples gave very consistent values of the signal efficiency and
background rejection, giving confidence
in the method. 

The total systematic uncertainty was obtained by adding the above 
contributions in quadrature, separately for the positive and negative deviations. 

\section{Results}
\label{sec-xsec}

\begin{figure}
\begin{center}
\includegraphics[height=10cm]{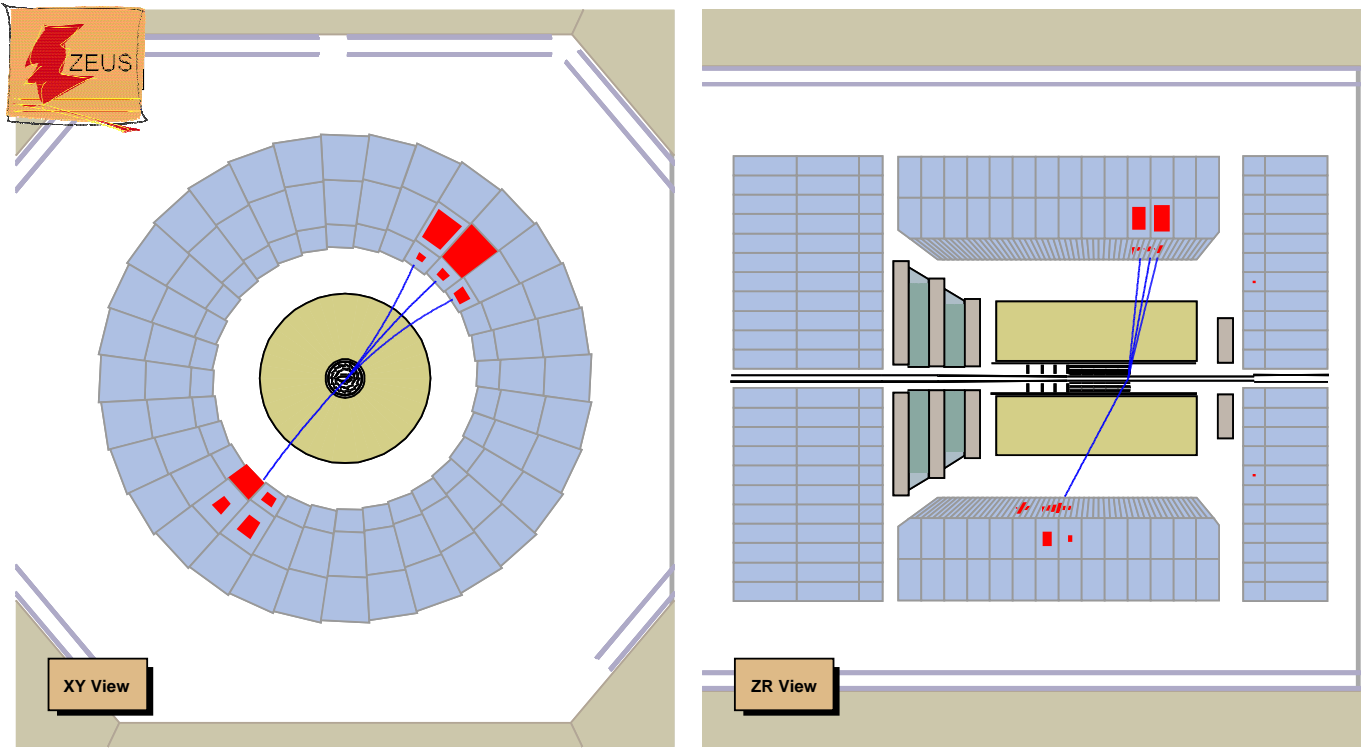}
\end{center}
\caption{
Event display of a data event selected in the jet-jet channel. 
One jet has one charged track, the other jet has
three charged tracks. The two
jets have $E_{T}^{\mathrm{jet1}}=7.8$ $\mathrm{GeV}$ and $E_{T}^{\mathrm{jet2}}=6.9$ $\mathrm{GeV}$
and the visible mass is $M_{\tau \tau}^\mathrm{visible}=15.6$ $\mathrm{GeV}$.} 
\label{fig-event}
\end{figure}

The total selected number of ditau candidates, $N_\mathrm{data}$, is 25, of which
13 were selected in $e^+p$  and 12 in $e^-p$ collision data, consistent with the respective
integrated luminosities.  
The total MC expectation was 
$34.8^{+3.9}_{-3.8}$ events, including an expected background, $N_\mathrm{bkg}$, of
$11.6\pm 3.9$ events (Table~\ref{tab-comparison}). 
The expected purity of the sample, evaluated from the {\sc Grape} ditau and the background 
MC, was $67\%$.    
The number of selected events in each channel is shown in Table~\ref{tab-comparison}.
One of the events in the jet-jet channel is shown in Fig.~\ref{fig-event}.

\begin{figure}
\begin{center}
\includegraphics[height=8cm]{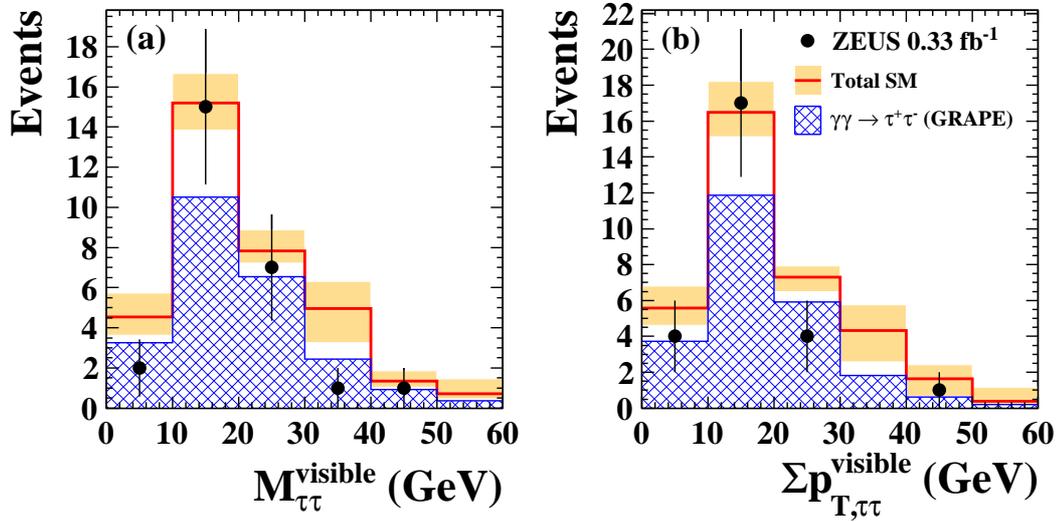}
\end{center}
\caption{Distributions of the events after all selection cuts 
as a function of (a)~the visible invariant mass of
 the tau pair, $M^\mathrm{visible}_{\tau\tau}$, and (b) the scalar sum of the 
transverse momenta of the 
two tau candidates, $\sum p_{T^,\tau \tau}^\mathrm{visible}$. The selected $e$-$e$-jet event
has two entries in the plots, one for each electron-jet combination. 
 The data (dots) are compared with the predictions of the
 sum of the  Monte Carlo expectations and to the ditau MC only. 
The shaded bands show the systematic 
uncertainty on the SM expectation.}
\label{fig-final}
\end{figure}

Figure~\ref{fig-final} shows 
the visible invariant mass, $M^\mathrm{visible}_{\tau\tau}$,
calculated from the two tau candidates, and the
 scalar sum of the visible transverse momenta of the two tau candidates
 in the event, $\sum p_{T,\tau\tau}^\mathrm{visible}$.
No event with a visible mass  $M^\mathrm{visible}_{\tau\tau}$ greater than
50 GeV was observed: 
the highest visible-mass candidate,
found in the $e$-$\mu$ topology, had $M^\mathrm{visible}_{\tau\tau}=49$ GeV.
 The MC predictions describe the data well and 
 no excess is observed in the high-mass or high-$\sum p_{T,\tau\tau}^\mathrm{visible}$ region. 
Only one of the 25 data events had three objects, corresponding to
an event with 
the scattered electron candidate in the detector, belonging to the $e$-$e$-jet topology.

The total cross section for ditau production was calculated 
for the kinematic region defined by 
$p_T^{\tau_{1,2}} > 5\,\gev$ and
$17^\circ < \theta^{\tau_{1,2}}<160^\circ$, where $p_T^{\tau}$ and 
$\theta^\tau$
refer to the transverse
momentum and polar angle, respectively, of the tau lepton. The cross section was
calculated as 
$$\sigma^\mathrm{kin}_{\tau\tau}=\frac{(N_\mathrm{data}-N_\mathrm{bkg})}{
                        \cal{A} \cdot \cal{L}},$$ 
where  $\cal L$ is the integrated 
luminosity of the data sample. The acceptance $\cal A$ was evaluated
from the {\sc Grape} ditau generator to be $1.23\%$. 
The total cross section was found to be
$$ \sigma^\mathrm{kin}_{\tau\tau}= 3.3 \pm 1.3 (\mathrm{stat.}) ^{+1.0}_{-0.7} (\mathrm{syst.}) ~\mathrm{pb}, $$
where the first uncertainty represents the statistical error and the
second the systematic uncertainty. 
The cross-section value is in reasonable agreement with  
 the SM expectation of $\sigma_{\tau \tau}^{\mathrm{SM}}=5.67 \pm 0.16~\mathrm{pb}$,
as evaluated from the {\sc Grape} MC generator.

\section{Conclusions}
\label{sec-conclusion}
 Events with two tau candidates with high transverse momentum 
 have been selected by the ZEUS experiment in the HERA II data 
and compared with the predictions of the Standard
 Model.
The tau leptons were identified through their decays into electrons, muons
or jets, with transverse momentum greater than 2 GeV (for an $e$ or a $\mu$) or
5\,GeV (for a jet).
The jet coming from the hadronic tau decay was identified with
a multi-variate discrimination technique employed to separate the signal from the 
QCD background. 
The selected events were dominated by the Bethe-Heitler $\gamma \gamma \to \tau^+ \tau^-$
process and  
the final-state topologies ($e$-)jet-jet, \eejet, \eemu
and \emujet  were considered. In total, 25 events were selected,  
compared to a MC expectation of
$34.8^{+3.9}_{-3.8}$ events, including 
$11.6 \pm 3.9$ events of expected background.
The distribution of events shows  good agreement with the Standard Model
expectation,
also at high values of the visible transverse momentum and visible invariant
mass of the tau pair. Therefore, no evidence of physics beyond the Standard Model is
found for tau-pair production.
The total cross section, in the kinematic region  
$p_T^{\tau_{1,2}} > 5~\gev$ and
$17^\circ < \theta^{\tau_{1,2}}<160^\circ$,  
was measured to be
$ \sigma^\mathrm{kin}_{\tau\tau}= 3.3 
\pm 1.3 (\mathrm{stat.}) ^{+1.0}_{-0.7} (\mathrm{syst.}) ~\mathrm{pb}. $

\section{Acknowledgments}

We appreciate the contributions to the construction and maintenance 
of the ZEUS detector of many people who are not listed as authors.
The HERA machine group and the DESY computing staff are especially acknowledged 
for their success in providing excellent operation of the collider and
data-analysis environment. We thank the DESY directorate for their strong support and
encouragement.

\vfill\eject

\providecommand{\etal}{et al.\xspace}
\providecommand{\coll}{Coll.\xspace}
\catcode`\@=11
\def\@bibitem#1{%
\ifmc@bstsupport
  \mc@iftail{#1}%
    {;\newline\ignorespaces}%
    {\ifmc@first\else.\fi\orig@bibitem{#1}}
  \mc@firstfalse
\else
  \mc@iftail{#1}%
    {\ignorespaces}%
    {\orig@bibitem{#1}}%
\fi}%
\catcode`\@=12

\tiny
\begin{sidewaystable}[htbp]
\begin{center}%
 \centering
  \begin{tabular}{|l||c|c|c|c|c|c|c|c|}
   \hline
   Topology & $e$-$\mu$ & $e$-$e$-$\mu$ & $e$-jet & $e$-$e$-jet &
              $\mu$-jet & $e$-$\mu$-jet & jet-jet & $e$-jet-jet \\
   \hline\hline
   Electrons & $N_{e}=1$ & $N_{e}=2$ & $N_{e}=1$ & $N_{e}=2$ &
               $N_{e}=0$ & $N_{e}=1$ & $N_{e}=0$ & $N_{e}=1$ \\
   \hline
   Muons & \multicolumn{2}{c|}{$N_{\mu}=1$} &
           \multicolumn{2}{c|}{$N_{\mu}=0$} &
           \multicolumn{2}{c|}{$N_{\mu}=1$} &
           \multicolumn{2}{c|}{$N_{\mu}=0$} \\
   \hline
   Tau jets & \multicolumn{2}{c|}{$N_\mathrm{jet}=0$} &
                 \multicolumn{4}{c|}{$N_\mathrm{jet}=1$} &
                 \multicolumn{2}{c|}{$N_\mathrm{jet}=2$} \\
   \hline
   $N_\mathrm{trk}$ & {$N_\mathrm{trk}=2$} & {$2\le N_\mathrm{trk}\le 3$} &
                      \multicolumn{6}{c|}{$2\le N_\mathrm{trk} \le 7$} \\
   \hline
   $(E-P_Z)$ & $< 45$~GeV & $< 60$~GeV & $< 45$~GeV &$< 60$~GeV  & $< 45$~GeV & $< 60$~GeV
                        & $< 45$~GeV & $< 60$~GeV \\
   \hline
   \multirow{5}{*}{Charge cuts} & $Q_e \neq Q_\mu$ & &
                                  $Q_e \neq Q_\mathrm{jet}$ & & & &
                                  $Q_{\mathrm{jet},1} \neq Q_{\mathrm{jet},2}$\,, &\\
    & & & $Q_\mathrm{jet}=\pm1$ & & & & $Q_\mathrm{jet}=\pm1$ & \\
    & & &                          & & & &                       & \\
    & $Q_e \neq Q_\mathrm{beam}$ & & $Q_e \neq Q_\mathrm{beam}$ & & & & & \\
    & (if $\theta_e > 1.0$) & & (if $\theta_e > 1.0$) & & & & & \\
   \hline
  \end{tabular}
 \caption
         {Definition and selection criteria for each event topology for 
         tau-pair production. The symbols $N_e$ and $N_\mu$ refer to the  
 number of selected electrons and muons in the event,  
$N_\mathrm{jet}$ indicates the  number of
tau-candidate jets. The 
 selection criteria and other variables are defined in the text.}
 \label{tab-cuts}
\end{center}%
\end{sidewaystable}
\newpage
\normalsize
\begin{table}[htb]
 \begin{center}
     {\bf ZEUS ditau events HERA II data (L=0.33~fb$^{-1}$)}
  \begin{tabular}{|c||c|c|c|c||c|} \hline
   Topology & ($e$-)$e$-$\mu$ & ($e$-)$e$-jet & ($e$-)$\mu$-jet &
   ($e$-)jet-jet & Total\\
   \hline\hline
   Data & 4 & 7 & 4 & 10 & 25\\
   \hline
   Total MC & $3.6^{+1.3}_{-0.3}$ & $8.8^{+1.8}_{-0.8}$ & $8.0^{+2.2}_{-1.2}$ & $14.4^{+2.2}_{-3.5}$ & $34.8^{+3.9}_{-3.8}$ \\
   $\tau^{+}\tau^{-}$ MC & $3.0^{+0.3}_{-0.2}$ & $5.3^{+0.3}_{-0.2}$ & $5.9^{+0.5}_{-0.5}$ & $9.0^{+0.4}_{-0.3}$ & $23.2^{+0.7}_{-0.7}$ \\
   \hline
  \end{tabular}
 \end{center}
  \caption{The observed and predicted ditau-event yields for
           the sum of the topologies and for
each channel separately. 
The total MC expectations include the sum due to ditau production, 
DIS neutral current 
interactions, photoproduction events and dielectron/dimuon pair production. 
The experimental systematic uncertainties are quoted on the MC expectations.
}
\label{tab-comparison}
\end{table}
\vfill\eject

\newpage

\newpage

\newpage

\vfill

\newpage

\vfill\eject

%
%
\end{document}